\documentclass[conference]{IEEEtran}
\IEEEoverridecommandlockouts
\usepackage{cite}
\usepackage{amsmath,amssymb,amsfonts}
\usepackage[numbers,sort&compress]{natbib}
\usepackage{dblfloatfix}
\usepackage{algorithmic}
\usepackage{graphicx}
\usepackage{multirow}
\usepackage{textcomp}
\usepackage{CJKutf8}
\usepackage{xcolor}
\usepackage{makecell}

\def\BibTeX{{\rm B\kern-.05em{\sc i\kern-.025em b}\kern-.08em
    T\kern-.1667em\lower.7ex\hbox{E}\kern-.125emX}}
\begin{document}

\title{SHECS: A Local Smart Hands-free Elderly Care Support System on Smart AR Glasses with AI Technology
}
% \author{Donghuo Zeng, Jianming Wu, Bo Yang, Tomohiro Obara, Akeri Okawa,\\ Nobuko Iino, Gen Hattori, Ryoichi Kawada, Yasuhiro Takishima}

% \author[1]{Donghuo Zeng}
% \author[1]{Jianming Wu}
% \author[1]{Tomohiro Obara}
% \author[1]{Akeri Okawa}

% \author[1]{Nobuko Iino}
% \author[1]{Gen Hattori}
% \author[1]{Ryoichi Kawada}
% \author[1]{Yasuhiro Takishima}

% \affil{KDDI Research, Inc., Saitama, Japan. \authorcr
%   \{\tt do-zeng, ji-wu, bo-yang, to-obara, ak-ookawa, \\ \tt no-iino, gen, ry-kawada, takisima\}@kddi-research.jp}

\author{\IEEEauthorblockN{Donghuo Zeng\IEEEauthorrefmark{1},
Jianming Wu\IEEEauthorrefmark{1}, Bo Yang, Tomohiro Obara, Akeri Okawa, \\Nobuko Iino, Gen Hattori, Ryoichi Kawada, and Yasuhiro Takishima
}

\IEEEauthorblockA{KDDI Research, Inc., Saitama, Japan.\\
\{\tt do-zeng, ji-wu, bo-yang, to-obara, ak-ookawa, \\ \tt no-iino, ge-hattori, ry-kawada, takisima\}@kddi-research.jp}}
\maketitle
\begingroup\renewcommand\thefootnote{*}
\footnotetext{Both authors contributed equally to this research}
\endgroup

\begin{abstract}
Some elderly care homes attempt to remedy the shortage of skilled caregivers and provide long-term care for the elderly residents, by enhancing the management of the care support system with the aid of smart devices such as mobile phones and tablets. Since mobile phones and tablets lack the flexibility required for laborious elderly care work, smart AR glasses have already been considered. Although lightweight smart AR devices with a transparent display are more convenient and responsive in an elderly care workplace, fetching data from the server through the Internet results in network congestion not to mention the limited display area. To devise portable smart AR devices that operate smoothly, we first present a non-intrusive, privacy-compliant, and no-keepalive-Internet required smart hands-free elderly care support system that employs smart glasses with facial recognition and text-to-speech synthesis technologies. Our support system utilizes automatic lightweight facial recognition to identify residents, and information about each resident in question can be obtained hands-free link with a local database. Moreover, a resident's information can be displayed immediately on just a portion of the AR smart glasses on the spot. Due to the limited size of the display area, it cannot show all the necessary information. We therefore exploit synthesized voice in the system to read out the elderly care-related information in its entirety. By using the support system, caregivers can gain an understanding of each resident’s condition immediately, instead of having to devote considerable time in advance in obtaining the complete information of all elderly residents, especially of new residents. Our experiments on this support system were conducted at an elderly care home in Tokyo. Our lightweight facial recognition model achieved high accuracy with fewer model parameters than current state-of-the-art methods. The validation rate of our facial recognition system in the practical use scenario was 99.3\% or higher with the false accept rate of 0.001, and caregivers rated the acceptability at 3.6 (5 levels) or higher.
\end{abstract}


\section{Introduction}
%Context
To facilitate the caregivers of the elderly care homes to provide the best care services to elderly residents, it is very important to always keep track of the elderly residents' information, such as the current health status of each resident and what kind of care they need. Currently, this health status is recorded manually by the caregivers. However, the elderly care industry suffers from staff shortages as the mobility of staff is very high compared to other industries. In addition, as the number of new elderly residents is tending to increase, there are limitations to how well optimal care services can be provided to individuals. For example, some new elderly residents are not equipped with ID tags for personal identification, and caregivers therefore need to expend considerable time acquiring the health and eating\&drinking status of new residents manually.

\begin{figure}[!t]
    \centering
    \includegraphics[width=0.5\textwidth]{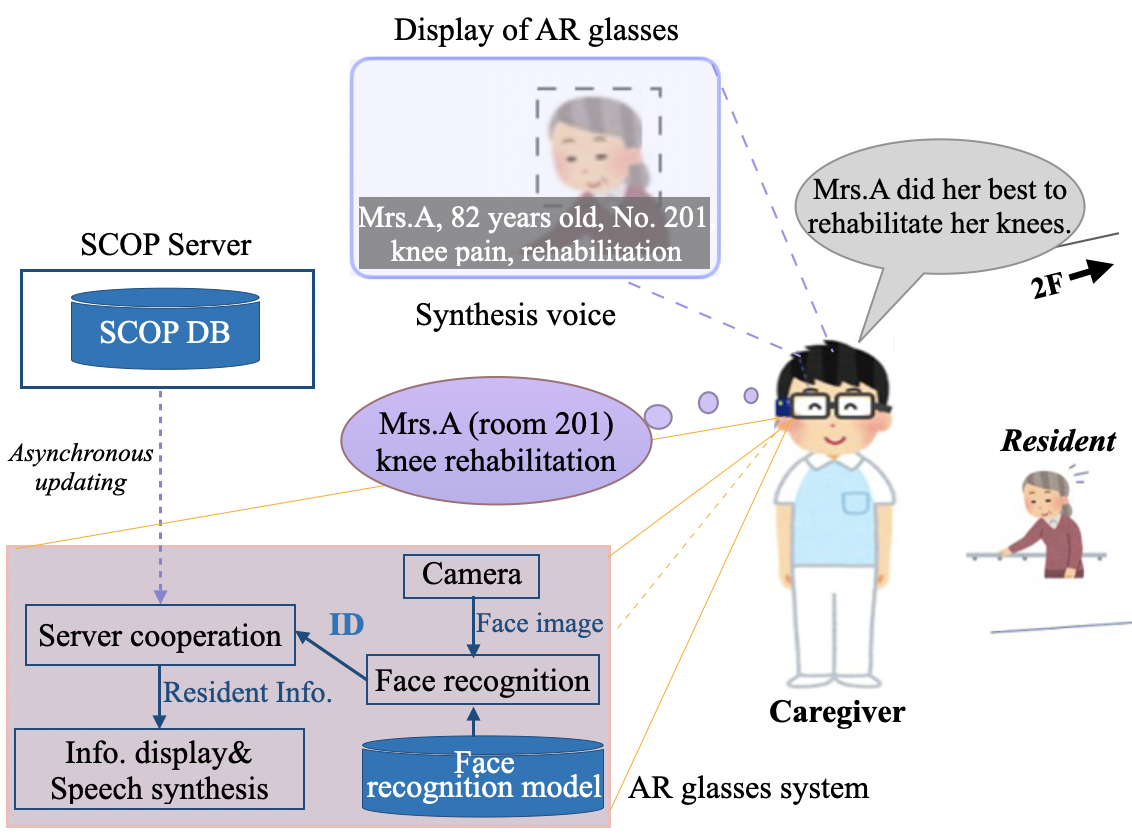}
    \caption{The Mechanism of our "Local Smart Hands-free Elderly Care Support System", where a rehabilitation room is located on the second floor.}
    \label{fig:overview}
\end{figure}
% Existing solution
% With the development of Augment Reality (AR) and AI technology, some elderly care support systems and related devices~\cite{klinker2020digital, kopetz2018smart, wuller2019scoping} are developed by some companies, such as hands-free elderly care devices that allows caregivers to quickly recognize the display information about residents to provide better services by presenting personal care status and giving responses to each resident. The use of this type of smart device will reduce the workload of caregivers and provide better care to the residents.
With the development of Artificial Intelligence (AI) technology, it is possible to reduce the caregiver workload and provide better care to residents through the use of AI. Some elderly care support systems with AI capabilities have been developed by a number of research teams, such as developing AI robots~\cite{mukai2010development, mukai2011tactile} to facilitate the elderly care work. 

% Challenges
The main challenge facing elderly care support systems is to quickly ascertain the condition of the elderly in the elderly care home and provide long-term care services smoothly by using information and communications technology (ICT). For example, the residents in our experimental elderly care home are unwilling to carry any ID tags or devices, and worry about the reliability of a robot, as they argue that this adversely affects their living experience and quality of life. In this case, it requires the caregivers to adopt and apply smart devices that help to maintain a normal life for residents.

% Gap
In the past, caregivers at the elderly care home at Zenkoukai~\footnote{https://www.zenkoukai.jp/japanese/}, a enterprise that operates and manages elderly care welfare facilities, could use smartphones and tablets to access residents' data on SCOP~\footnote{https://bi.biopapyrus.jp/db/scop.html}, which residents' health data such as the body temperature and blood pressure are centrally managed. However, in the field of elderly care, there are many tasks that require the use of both hands, such as bathing. Sometimes, there are cases where a rapid response is required, such as when a resident has a fall. In these cases, having to use the hands to operate a smartphone or tablet is a hindrance.
% \In addition, some AR devices with face recognition capability to help caregivers identify residents to obtain their healthcare information through Internet or Bluetooth. However, data found by the system, such as the face photo that only matches the registered frontal image. If you register a skewed face image, it will cost much time to look for it, but then even if you want to recognize a single face, you need to register a lot of different face orientation images, such as right, left, up, down, and etc. toward face image.

When caregivers are communicating with a resident face-to-face and needs to ascertain the resident's condition immediately through the care support system, speed is of the essence in improving care. However, in the extreme case, these 400 residents in one building fetch data from remote servers synchronously can cause an increase in data transmission levels, resulting in network delay and congestion. 
Moreover, to make the device easier to use when the caregiver is in physical contact with the elderly person's body, we hope to eliminate cables as these can get tangled, tripping over the cable may result in damage, and it may make the device uncomfortable to wear. In this case, we need to develop a hands-free care support system by using AR smart glasses to recognize the face of the resident, and which is also capable of accessing the local server that manages the resident’s healthcare data, and present this information on the device for caregivers.

% Proposal
In this paper, we introduce a SHECS system executed on the smart glasses. After installing the system on smart glasses, the device conducts a two-stage face recognition to achieve person identification including face detection and face recognition. 
%methodology
Our face recognition system benefits from a lightweight network that saves computational resources. Moreover, the system is capable of correcting various orientations such as face skewed to the side or presented front on to improve the system's performance. Once the ID of somebody is obtained, the system will access the local server inside the device (which is updated asynchronously from the SCOP server) to acquire the resident's relevant information in a short period of time. Beside, it was important to reduce the load of the AI's operation and the computational time. For example, by using AI technology to automatically eliminate useless information from data when extracting features from mugshots, and by exploiting a fast matching algorithm. In the end, our lightweight facial recognition model maintains high accuracy with fewer model parameters than current state-of-the-art methods, and our system was able to identify 10,000 mugshots in 1 second and 100 mugshots in 0.1 seconds on the smart glasses.

Furthermore, the display screen on the smart glasses is too narrow to present the resident's full information. 
Accordingly, we exploit a Japanese text-to-speech technology to instantly read the information in its entirety out using N2 Speech Synthesis Software~\footnote{https://www.kddi-research.jp/english/products/n2.html}. The advantages of this technology are that it employs a lightweight model and requires little memory.

% Result and implication
Following the results obtained from our preliminary experiments, we perform a real-world experiment at actual long-term care facilities with accuracy and acceptability to both caregivers and residents used as metrics. We achieved high recognition accuracy regardless of the orientation of the face, without increasing the amount of computation.
%Survey and selection of AR smart glasses Prototype of an application that links the face recognition technology with the resident DB. 
% In the evaluation of recognition in the actual environment of long-term care facilities and evaluation of acceptability to caregivers and residents. The accuracy of facial recognition achieved 99.9\% or higher, and the acceptability of caregivers reach 3.5 (5 levels) or higher.

The remainder of this paper is structured as follows. Section II introduces related work on smart care support systems and technologies related to our system. Section III presents the architecture of our model and Section IV reports the experimental results and analysis. Finally, in Section VI, we present our conclusions.

\section{Related work}
\subsection{AI for Smart Care System}
Now more than ever, a high-quality smart care system depends on the computational power when interacting with the vast, unstructured, and ever-changing clinical information or resident condition. AI brings a promising direction for the practice of care system because AI has stupendous potential to develop productively care system and provide huge opportunities. The review paper on application scenarios~\cite{seibert2020rapid} was published in 2020, they selected 699 full papers and 6,818 other texts and found that there are few publications for elderly care (nursing) homes and home-care related.

In the past, to improve the care support system in elderly facilities, this study~\cite{watanabe2009design} proposed an elderly care support system that can be applied to information storage devices to record information about all patients by implementing a domain name system. 
\cite{chen2015development} proposed a AR-based surgical navigation system to reduce the risk and improve the chances of successful surgery by using a transparent display worn on the head. %
This study~\cite{mitrasinovic2015clinical} presented online smart AR glasses that can display patient information and record a video of the patient, with the goal of improving patient care and reducing healthcare or nursing costs.
Another study~\cite{ruminski2015interactions} investigated how health information system data can be matched with the corresponding patient using smart glasses. 

With the use of AI technology in the smart care systems, more and more relevant works are emerging, \cite{clancy2020artificial} discusses the methods of AI and give some examples of how AI can help healthcare, especially nursing.
In this paper~\cite{grayson2020dynamic}, an AI system was presented that helps blind people detect the location, gaze direction, and identity of a nearby person. When wearing the device, the user can receive the voice and visual information of other people, which can help them communicate with other people who are also healthy.

With the great success of AI technology like computer vision~\cite{yu2018category, zeng2020deep, yu2019deep}, robots have also become very helpful assistants for long-term care~\cite{watson2020rise}. For example, AI robots can reduce the workload of caregivers and improve patient-caregiver interaction. 
The RIBA robot~\cite{mukai2010development, mukai2011tactile} has human-like arms and is designed to complete heavy physical works and replace human workers. 
For example, the RIBA robot can carry a patient from a couch to a wheelchair and back again, and also work with caregivers. 
To learn more about the robot's potential as a real-world care assistant, \cite{carros2020exploring} launched a 10-week study was conducted in a nursing home using the Pepper robotic platform, with residents and caregivers among the participants. They found out the facilitating and hindering factors of the robot and were able to better understand the supportive physical activation, cognitive training and social facilitation in human-robot interaction with the elderly. HomeMate~\cite{lee2020toward} is the next generation of care robots for the elderly, they develop five service scenarios by optimizing synergy with an ecosystem of care robots for the elderly.

\subsection{Facial Recognition and Speech Synthesis}
\subsubsection{Facial recognition}
Conducting facial recognition research to help the elderly is a long-standing research.  Robot companion and caregiver for elderly~\cite{sharkey2011children} is a taking hold idea, face recognition is the main factor of such robots that can recognise the elderly face then take next action. Nursebot is another different kinds of robot that involves multi-disciplinary technology, for example, the Pear Nursebot~\cite{pollack2002pearl} is mobile robotic assistant for elderly including recognition function to identity the elderly people then telling them the routine activities or guiding them through their environment. Healthcare systems for the safety of elderly people at home~\cite{rougier2006monocular} is to solve the social problem in the growing population of elderly, the face recognition and other technique here can help elderly to track their behavior such as detecting falls of elderly. With the development of face recognition technologies~\cite{andrejevic2020facial, mundial2020towards, manssor2019tirfacenet, schroff2015facenet}, it is not limited in the above applications for elderly. Recent years, face recognition can be applied in many fields. For example, the work~\cite{park2019deep} developed a smart door lock system with face recognition to help elderly open the door without any additional device such as smartcard, keys or smartphone. Real-world face recognition systems~\cite{mandal2016face} can help elderly through deploying IP cameras in the elderly care center in Singapore, such system have 4 functions including the time of a resident in toilet, the location of a resident, who left the emergency room, and who left center. Long-Term Health Monitoring system for elderly with the combination of facial recognition~\cite{luo2018computer} and non-intrusive, privacy-compliant wearable sensors is very useful technology to track their activities.

\subsubsection{Speech synthesis}
Automatic personal synthetic voice construction~\cite{bunnell2005automatic} can be achieved via two types of software: InvTool and BCC. InvTool~\footnote{http://www.asel.udel.edu/speech/InvTutor/index.htm} enables the user to create his/her own corpus of speech suitable for a speech synthesis application. BCC is the other program. It can compile the speech corpus recorded with InvTool into a set of files. The study of turn-taking in conversations includes the issue of “additive effect”. ~\cite{hjalmarsson2011additive} examined this effect by using the synthetic voice to leverage the generated turn-taking cues and found that in dialogues natural human speech could be replaced by the synthetic voice. ~\cite{damacharla2019effects} presented a voice-based synthetic assistant to facilitate the training of medical first-responders to improve efficiency during an emergency. This research~\cite{paris1995effect} further exploited the different impacts of synthetic voice by inviting sixty people to listen to ten other people in three different modes: natural speech, and low or high intelligibility. The results showed that different modes affect the quality of synthetic voices. A new CNN-LSTM structure, called the objective prediction model~\cite{mittag2021deep} for synthetic voice, was used to evaluate the quality of synthetic voice in different situations. They show that the reliability of deep learning-based naturalness prediction can be improved by transfer learning from speech quality prediction models that are trained on objective POLQA scores. 

Facial recognition technology is expected to be used in elderly care homes. To our knowledge, there are no existing AR glass-based systems for elderly care. On the other hand, iFalcon Face Control\footnote{https://nntc.digital/projects/ifalcon-face-control-mobile-facial-recognition-for-smart-glasses/} proposed an smart AR glasses solution for facial recognition, in which the face data is transferred from the smart AR glasses to a portable computer and the matching result retrieved from the computer via Bluetooth. However, such a solution cannot avoid the network delays and congestion in an elderly care home with hundreds of residents.  

\section{Architecture}
We apply the SCOP server to manage residents’ healthcare data, and recognize the faces of the residents through the smart AR glasses based on photos taken beforehand from the front or from an oblique angle. The residents’ healthcare information will be projected onto the lens of the AR glass, at the same time, the system will read out the resident’s information by audibly providing the information through a bone conduction earphone~\footnote{https://aftershokz.jp/products/aeropex}. The content of the voice can be set in various ways depending on the situation. This section consists of three subsection including facial recognition technology to identify the resident, the content displayed on the AR glasses, and speech synthesis.

\begin{figure*}[h]
    \centering
    \includegraphics[width=1.0\textwidth]{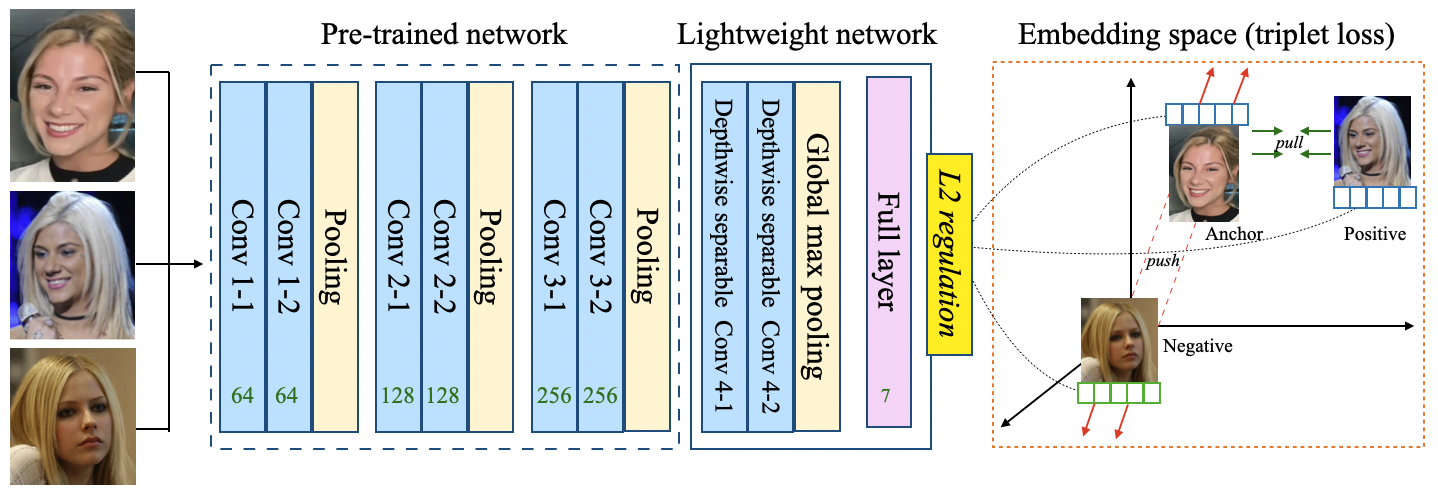}
    \caption{Overview of our facial recognition architecture, comprising three parts: pre-trained network, lightweight network, and embedding space. By combining the middle layer features of the learned model with the lightweight network, we could apply a lightweight learning model that significantly reduces computational resources. We employ triplet loss to update the parameters of the whole deep neural network and output the final representation of each input in the embedding space, where the “Anchor” image and the “Positive” image are pulled closer together, while at the same time, pushing the “Anchor” image and “Negative” image further apart.}
    \label{fig:model}
\end{figure*}

\subsection{Facial Recognition}
\begin{figure}[h]
    \centering
    \includegraphics[width=.48\textwidth]{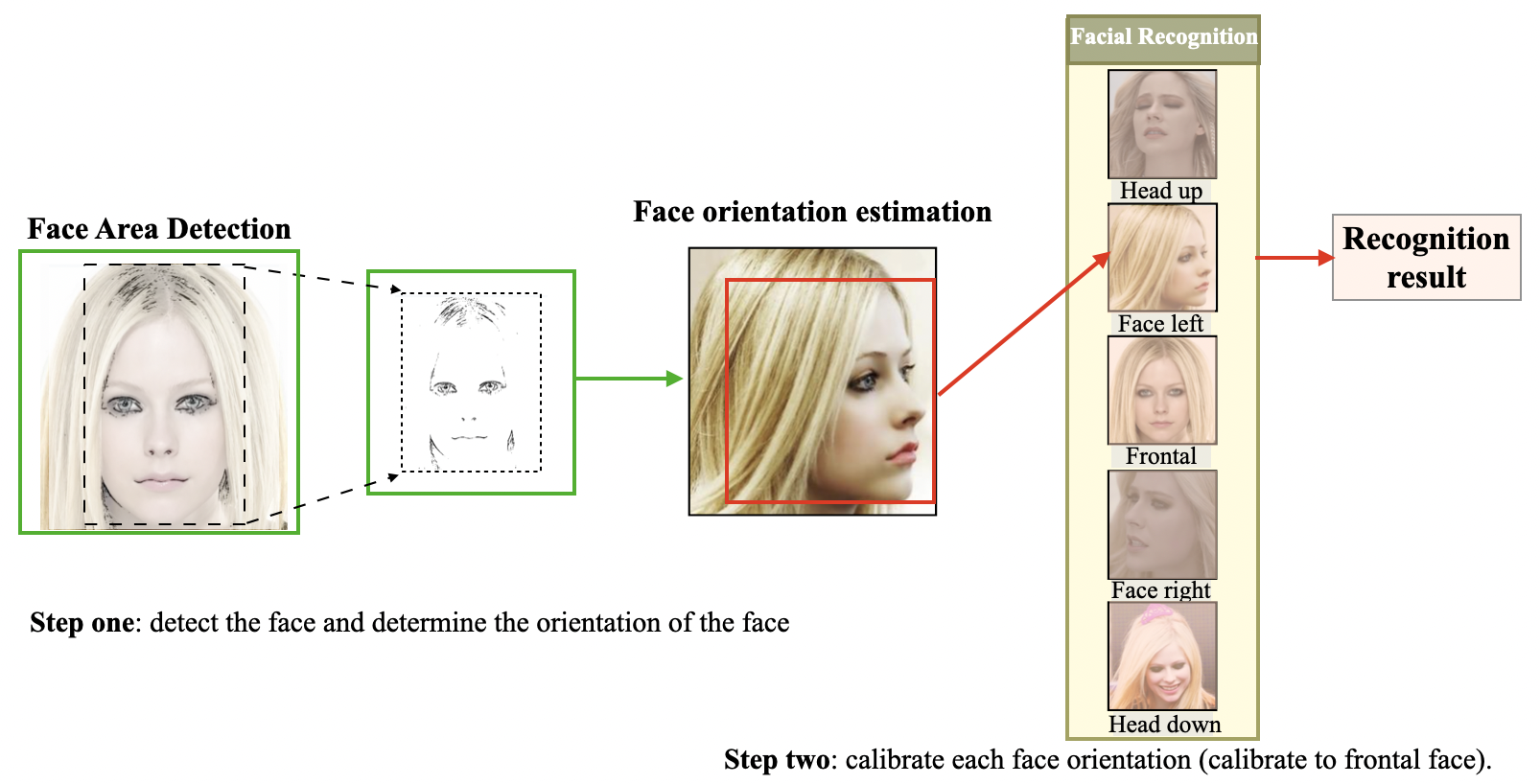}
    \caption{The process of calibration for each face orientation includes two steps: Detect facial features and match from known faces}
    \label{fig:orient}
\end{figure}

To quickly ascertain someone’s identity by facial recognition technology requires a recognition system that can be integrated with the work process of caregivers when the caregivers need to check the condition of residents. In this case, the existing high-level accuracy deep neural network model needs to be modified by reducing the number of parameters. Inspired by~\cite{wang2019lightweight}, we extract the discriminative representations of identities from the middle layer of a high accuracy DNN model, then feed these representations into a pre-trained network (VGG-retrain-v1), which comprises several lightweight CNN-based layers and global max pooling as shown in Fig.~\ref{fig:model}.

To implement the lightweight facial recognition system and enable the system to be applied in a small Internet-of-Things (IoT) device, we design the entire architecture for facial recognition incorporating three main components:
1)the learned network consists of three CNN-based blocks of a pre-trained network, and the pre-trained network refers to the CNN-B~\cite{simonyan2014very} that includes eight convolution layers, four pooling layers, and two fully connected layers. 
2)the lightweight network contains two CNN-like blocks in order to reduce the number of parameters. Compared with the convolutional layers of the pre-trained network~\cite{Sun_2015_CVPR}, we apply a depthwise separable convolutional layer as the lightweight layers to reduce the computational cost by separating the depthwise convolutional layer and the point-wise convolutional layer. As a consequence, the lightweight layers can reduce the number of parameters by about 90\%.
3) the embedding space is to apply triplet loss to directly reflect what kind of features we want in the facial recognition training network. In particular, we select one facial image as an anchor, all other images with the same identity are marked as positive, with different identities being marked as negative.

As a result, we were able to provide a native library on Android and achieved identification performance of 10,000 people per second on the android smartphone and the AR glass.
In general, the facial recognition system has two steps: 
step 1. the facial recognition model retrieves facial features from the photograph of a face, step 2. it compares the facial features with a database of known faces to find a match. The more registered known faces there are in the database, the greater the CPU search time and the more memory is consumed. Therefore, we implemented a native-c based search program for 2-step processing. In step 1, the face recognition model converts the facial features of each elderly resident into 128 bytes and allocates its ID into 4 bytes (unsigned int). This means that each elderly resident only needs 132 bytes for its registration. Then in step 2, we concatenate all the registration data of the elderly residents into one binary file and implement a binary search with pointers in C language to enable fast execution.

In addition, we need to consider the various face orientations in order to improve the effective features more efficiently. In cases where the orientation angle of the face in the electronic data is not always a frontal face view, facial orientation recognition~\cite{gupta2017robust} is important to improve the accuracy of the machine perception system. In this paper, we calibrate each non-frontal face orientation with a mixture of data which contain various face orientations. The process can be seen in Fig.~\ref{fig:orient}. The first step is to detect the face and determine the orientation of the face, the second step is to calibrate each face orientation to the frontal face view using an external library~\cite{zhang2016joint}.

\subsection{Display Content}
\begin{figure*}[h]
    \centering
    \includegraphics[width=1.0\textwidth]{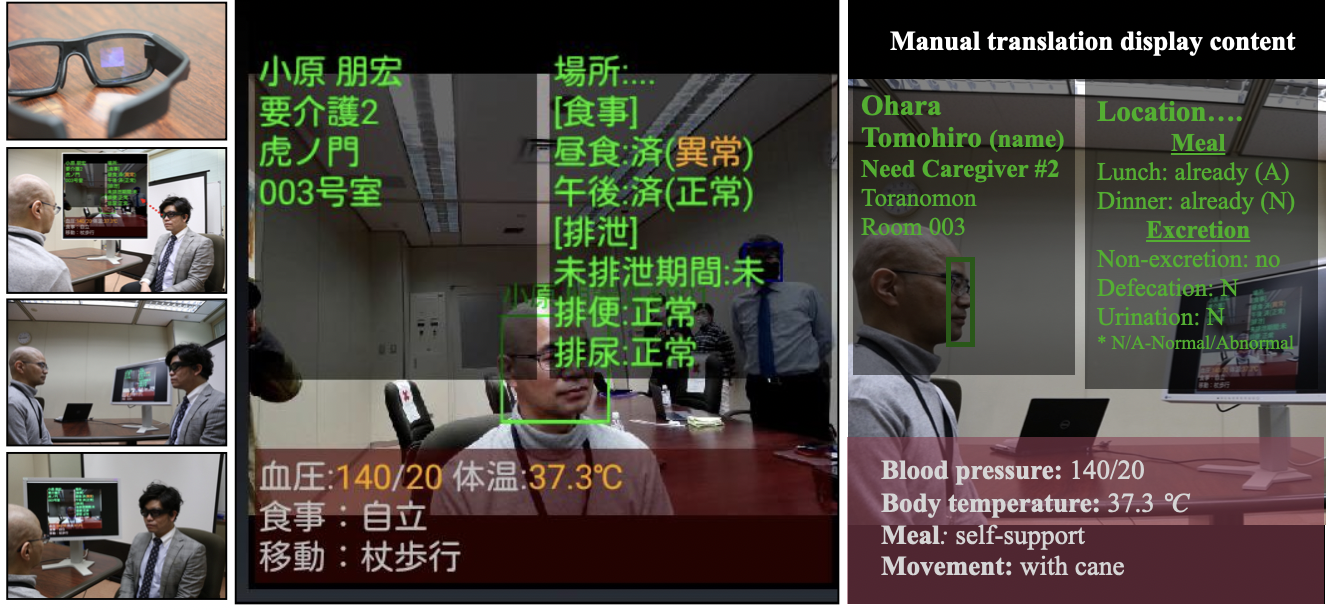}
    \caption{The Demonstration of the "Smart Hands-free Elderly Care Support System" and Collage of Images}
    \label{fig:display}
\end{figure*}
When the caregiver wears the AR glasses, the camera will capture the resident’s face, and the resident’s healthcare information will be displayed on the right-hand lens of the glasses, as seen in Fig.~\ref{fig:display}. The resolution of the AR glass is 1920x1080 pixels. The displayed information includes “resident’s name”, “level of care required”, “elderly care home name and room number”, “blood pressure”, “body temperature”, “Meal status”, “degree of mobility”, “eaten or not”, and “excretion or not" and "local time”.

%In addition, the resident's name will be audibly played from the earphone.
The benefit of displaying this content on the glasses is that the caregiver can take action immediately according to the displayed content. For example, if the caregiver notices the resident’s blood pressure is abnormal, he/she can make an informed decision. If the resident currently has high blood pressure, the bathing time should be reduced. If the resident has not passed urine or feces, the caregiver would ask the resident if he/she needs to go to the bathroom. In addition, even if the caregiver has not yet obtained detailed information such as for a new resident, the caregiver can easily take appropriate measures because the new resident's information and his/her current physical condition can be displayed by the devices.

\subsection{Speech Synthesis}
After the AR smart glasses recognize the face of a resident, the smart glasses will read the information out in Japanese using speech synthesis technology, as seen in the Fig.~\ref{fig:overview}. We apply a lightweight text-to-speech technique on a standalone micro-controller board for Internet of Things (IoT), which is a memory-saving and lightweight operation. In detail, the input of the speech synthesis system is data in text form, and the system will generate the intermediate data with the help of an initial word dictionary. By using the hidden Markov model (HMM)-based text-to-speech method, the system will generate a lightweight model and reduce the number of parameters. Finally, synthetic speech will be generated by multi-band signals system (MBSS) technology~\cite{alim2013enhancement} that uses the multiband sine wave synthesis method.

Furthermore, it is possible to freely add new words to the word dictionary that are not in the current word dictionary. Homonyms such as people’s names can be read out properly by assigning reading kana, for example, the Japanese name Kohara (\begin{CJK}{UTF8}{min}コハラ\end{CJK}Kohara, \begin{CJK}{UTF8}{min}オハラ\end{CJK}Ohara, \begin{CJK}{UTF8}{min}オバラ\end{CJK}Obara).

\section{Experiments}
\subsection{Evaluation Metric}
To evaluate our smart hands-free elderly care support system, we designed some evaluation metrics to leverage and investigate the acceptability of the devices while we conducted the experiments.

In the preliminary experiments on the premises, the evaluation was focused on the accuracy and robustness of face recognition against differences in face orientation and illumination of subjects, which included official full-time caregivers and temporary employees. The evaluation conditions are divided into three levels with a range of variations in face orientation. Finally, we evaluated our facial recognition methods under a combination of two evaluation conditions. Some examples of these combinations of evaluation conditions are as follows.
\begin{itemize}
    \item Case one: bright environment, the orientation of the face on the AR Smart Glasses screen varies within a range of 60° from the normal line of the screen.
    \item  Case two: normal environment, the orientation of the face on the AR Smart Glasses screen varies within a range of 40° from the normal line of the screen.
    \item  Case three: dark environment, the orientation of the face on the AR Smart Glasses screen varies within a range of 20° from the normal line of the screen.
\end{itemize}

There are nine cases in total according to the combination of evaluation conditions, 20°, 40°, and 60° for face orientation, and bright, normal, and dark for illumination. In our paper, we will show the result of face orientation be set as 0° and 45° in the normal environment. Beside, the distance between caregiver and resident is also important, in our experiments, we set the similar distance ranges from 50 cm to 100 cm.

\begin{figure}[!t]
    \centering
    \includegraphics[width=.5\textwidth]{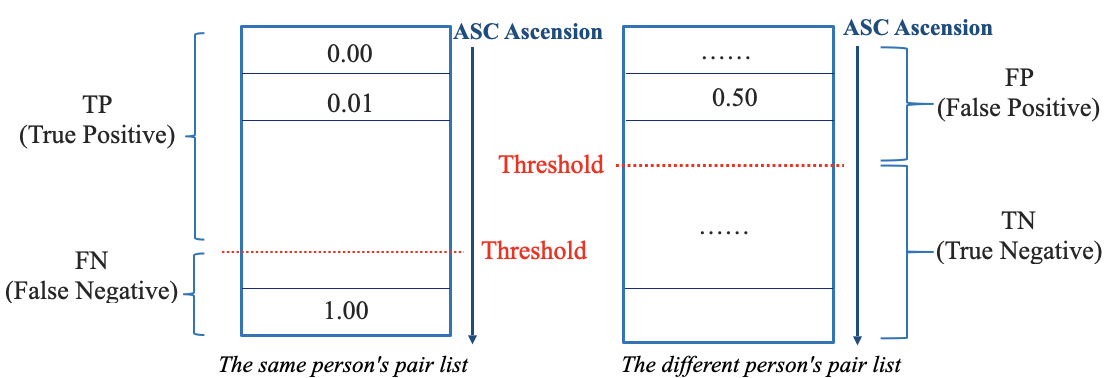}
    \caption{The accuracy of facial recognition is according to the validation rate for the same person and the false accept rate for a different person, where the validation rate is where the same people are recognized, while false accept rate is where different people are perceived as being the same person.}
    \label{fig:metrics}
\end{figure}
In addition, we tested acceptability as another evaluation in the production experiment at the Zenkoukai site in 2020/10/13 to 2020/12/18. The evaluation items include:1) Facial recognition accuracy. 2) Display content on AR smart glasses. 3) Content to be read to earphones.
4) Time taken to display. 5) Time taken to read out. 6) Do you actually want caregivers to use it. 7) Impressions on residents of caregiver wearing this device. The AR smart glasses, chargers, earphones, and etc. will be discussed separately from the viewpoint of handling personal information. And we also investigated the impression of caregivers using a 5-point evaluation based on the mean opinion score (MOS). The product was evaluated on a 5-point scale of [very good], [good], [normal], [bad], and [very bad].

As for the accuracy of facial recognition, refer to~\cite{schroff2015facenet}, we define the percentage of correctly recognized face images as the validation rate, while the percentage of incorrectly recognized face images as the false accept rate, as seen in the Eq.(\ref{eq:rate}). In our experiments, the validation rate for the same person and the false accept rate for a different person are the main evaluation indicators. There is a trade-off between the validation rate and the false accept rate, as shown in Fig.~\ref{fig:metrics}, where the threshold is determined by the false positive (FP) or false accept rate that is set in advance. For example, if the false accept rate is 0.001, the threshold decides the validation rate should be $1-threshold=99.9\%$. If Val@FAR=0.001 is the rate that faces are successfully accepted (TP/(TP+FN)) when the rate that faces are incorrectly accepted (FP/(TN+FP)) is 0.001. Overall, a good facial recognition system requires both a high validation rate and a low false accept rate.
\begin{equation}
    \begin{aligned}
        Validation\, Rate (VAL) = TP/(TP+FN) \\
        False\, Accept\, RATE (FAR) = FP/(FP+TN)
        \label{eq:rate}
    \end{aligned}
\end{equation}
%Calibration of face orientation, optimization of facial feature learning models, etc.: Study of smart glasses operation methods

\subsection{Setting}
We utilize Google’s FaceNet algorithm~\cite{schroff2015facenet} as the baseline of our proposed approach. We map each face image to a compact Euclidean space using a deep convolutional network. If the Euclidean distance of two images is close, this indicates they belong to the same person. During iteratively training the model, we created a triplet that consists of Anchor, Positive, and Negative to compute the cost function, then applied triplet loss as an objective function to update the parameters of our deep neural networks. The preliminary experiment that investigated the execution speed was performed on an Android device released in 2016 and our AR device, the detail is seen in Table.~\ref{tab:config}.
\begin{table}
\centering
  \caption{The configuration parameters of the device we applied.}
  \label{tab:config}
  \begin{tabular}{c|c} 
  \hline
    \thead{Parameter name}&\thead{Value}\\ \hline
    Operating System & Android OS \\
    Processor&ARM Cortex-A53 \\ 
    Companion APP & Android 6.0 or lateriOS 11.0 or later\\ 
    Memory & 5.91 GB (expandable via microSD card)\\ 
    Camera & 8MP with 720p video\\
    Battery & 470mAh Lithium polymer \\
    \hline
\end{tabular}
\end{table}

\subsection{Results}
\begin{table*}[h]
\centering
  \caption{The comparative evaluation of open sets including the recognition accuracy and execution speed}
  \label{tab:eval}
\begin{tabular}{c|cccccc}
\hline
    \thead{Evaluation \\ conditions}&
    \thead{Our model+ \\ VGGFace2~\cite{cao2018vggface2}}&
    \thead{Our model+ \\ CASIA-WebFace~\cite{yi2014learning}}&
    \thead{Google Facenet~\cite{schroff2015facenet}\\+VGGFace2}&
    \thead{Google Facenet~\cite{schroff2015facenet}\\+CASIA-WebFace}&
    \thead{1.00Fast-FaceNet \\VGGFace2~\cite{xu2020lightweight}} &
    \thead{0.25Fast-FaceNet \\VGGFace2~\cite{xu2020lightweight}}\\
\hline
    LFW pair.txt	&
    \makecell{\underline{Accuracy: 0.99680} \\ \underline{VAL 0.98633}@ \\ \underline{FAR=0.00100}} &
    \makecell{Accuracy: 0.99620 \\ VAL: 0.98833@ \\ FAR=0.00100} &
    \makecell{Accuracy: 0.99630 \\ VAL: 0.98600@ \\ FAR=0.00100} &
    \makecell{Accuracy: 0.99550 \\ VAL: 0.98357@ \\ FAR=0.00100} &
    \makecell{Accuracy: 0.98630 \\ VAL: 0.97452@ \\ FAR=0.00100} &
    \makecell{Accuracy: 0.79530 \\ VAL: 0.77910@ \\ FAR=0.00100}\\
    Params& \underline{2.4m}&	2.4m&	7.5m&	7.5m&	4.8m&	0.9m\\ 
    100 people& \underline{86ms}&	86ms&	632ms&	632ms& 196ms&	63ms\\ 
    1000 people& \underline{270ms}& 270ms&   System Halted&  System Halted& 671ms& 206ms \\
    10000 people& \underline{925ms}& 925ms&	System Halted&	System Halted& 1487ms&	704ms\\
\hline
\end{tabular}
\end{table*}
In our experiment, we apply the LFW (pair.txt file)~\cite{huang2008labeled} dataset to evaluate our system, since it will consume too much time if we use all the cross-pairs of the LFW dataset. Accordingly, we select the same 6,000 pairs as~\cite{huang2008labeled} from the LFW dataset as the evaluation data. The results of a comparative evaluation of open sets by evaluating the recognition accuracy and execution speed are shown in the Table.~\ref{tab:eval}. Overall, our system achieves accuracy of 99.6\% or higher and is superior in terms of both recognition accuracy and execution speed. In particular, our model with VGGFace2 or CASIA-WebFace can achieve the best accuracy and VAL/FAR in the experiment. An accuracy of 99.68\% and 99.62\%, with a VAL of 98.63\% and 98.83\%, and a FAR of 0.1\% and 0.1\%, respectively, were achieved. We also assess the execute time in relation to the change in the number of people, the execute time includes the facial feature retrieval cost and the face matching cost. We select 100, 1,000, and 10,000 people to be recognized. Since most of the existing methods were implemented only by Java- or Python-based functions for matching faces, we apply our C-based fast matching algorithm to all methods to provide a fair comparison of execution speed. The result proves that our lightweight network structure maintains high accuracy with fewer model parameters than the current state-of-the-art methods, which can reduce the time cost of face recognition and search time, and as the number of people increases, the time saving becomes more obvious. On the other hand, the accuracy of the 0.25Fast-FaceNet model was only 0.7953, which does not meet the requirements, even though it has fewer parameters.

%facial recognition cost and the resident information retrieval cost   //keep it by swordwu

\begin{table}
\centering
  \caption{The ACCURACY OF FACE RECOGNITION IN AR SYSTEM WITH CHANGES IN VIEWING ANGLE.}
  \label{tab:freq}
  \begin{tabular}{c|c|cc}
  \hline
  \thead{Viewing angle}&\thead{Image pairs}&\thead{VAL}& \thead{FAR}\\ \hline
  0-degree (front) &6000 & 99.9\% & 0.00100\\
  45-degree &6000 & 99.3\% & 0.00100 \\
  \hline
\end{tabular}
\end{table}

The face orientation and illumination of the image will influence the accuracy, VAL, and FAR of the system. By considering the limitations of the experimental environment, in this paper, illumination was set as normal light in all experiments, and the face orientation was changed between frontal face and face at a 45-degree angle. From the Table.~\ref{tab:freq}, we have obtained excellent results for VAL and FAR using face recognition AR technology regardless of the changes in viewing angle. We use different 6000 image pairs (the same standard as LFW) for two different viewing angles.
In the end, when the images are front face, the VAL is 99.9\% and the FAR is 0.001\%, while the VAL is 99.3\% and the FAR is 0.001\% where the facial image is at a 45-degree viewing angle.

Since the developed recognition system needs to be implemented on smart glasses to assist the caregivers, apart from the requirement for high-level accuracy, it is also very important to investigate the system’s acceptability to caregivers in relation to several factors. We administered a questionnaire to caregivers about acceptability when they used the smart glasses. They were asked to fill out the questionnaire by selecting one of five pre-defined ratings. The rating scale is as follows, 5 (Strongly agree), 4 (agree), 3 (undecided), 2(disagree), and 1 (completely disagree). The questions of the questionnaire were as follows: (A) Did the displayed contents match the residents? (B) Are you satisfied with the content displayed on the right lens of the smart AR glasses? (C) Is there is any additional information you would like to see on the smart AR glasses? (D) Is there any information displayed on the smart AR glasses you feel is unnecessary? (E) If the above questions do not cover all your concerns, please give us specific examples. For example, the display takes too long to appear, or it is hard to identify whose information is being displayed, etc.

% Since the development recognition system needs to apply to the smart glasses to assist the caregivers, except for the requirement of high-level accuracy of the system, it is also very important to investigate the acceptability of objects, including caregivers and elderly residents. We launch a questionnaire about the acceptability when using smart glasses by caregivers. we invite them to fill out the questionnaire by selecting the pre-defined ratings. The rating scale is as follows, 5 (Strongly agree), 4 (agree), 3 (undecided), 2(disagree), and 1 (completely disagree). The questions of the questionnaire include:
% (A) Did the displayed contents match the residents?
% (B) Are you satisfied with the content displayed on the right eye of the AR glasses?
% (C) If there is any additional information you would like to see on your AR glasses.
% (D) If there is any information that you feel is unnecessary in the displayed contents of AR glasses.
% (E) If you are not satisfied with the above question, please tell us the specific reason. Example. The display is too late, too early to know who's information, etc.

\begin{table}
\centering
  \caption{A Caregivers' Responses to QUESTIONNAIRE concerning their feeling about the AR Device.}
  \label{tab:staff}
  \begin{tabular}{c|c}
    \hline
    \thead{Feeling} &  \thead{Number of People}\\ \hline
        Scared & 5 \\ 
        Funny looking & 9 \\ 
        Especially Scared & 2\\ \hline
\end{tabular}
\end{table}

\begin{table}[!t]
\centering
  \caption{A caregivers' responses to QUESTIONNAIRE on acceptability.}
  \label{tab:staff-resident}
  \begin{tabular}{c|cc}
 \hline
    \thead{Participant's Role} &\thead{Number of People} &\thead{Average Rate Score}\\ \hline
        Caregivers &  6 &3.6\\ \hline
\end{tabular}
\end{table}
% Moreover, from the staff questionnaire, we received some concerns about the weight, battery life, and heat.

According to the answers from six caregivers when asked about using the smart glasses, some of the satisfied caregivers expressed the following: “I appreciate it because I don’t have to search for information on using my cellphone”, or “It is intuitive and easy to use, even for new caregivers who have never used it before”, or “The displayed content is easier to read than I expected”. But some caregivers expressed the concern that “the smart AR glasses were heavy”, or “they look like sunglasses”. 

% shown in the Fig.~\ref{fig:question}, the results are shown in the Table.~\ref{tab:staff}. Five people felt scared, nine people felt the glasses looked funny, and two people commented that they were especially scared and wanted the caregivers to take off the glasses immediately.

% According to the reflection of six caregivers when using the smart glasses, some of the satisfied caregivers feel: "I appreciate it because I don't have to search for information on my cellphone", or "It is intuitive and easy to use, even for new caregivers who never use it before", or "The displayed content is easier to read than I expected". But some caregivers will worry that "the AR glasses were heavy", or "it looks like sunglasses". 

% % residents
% As for the residents' questionnaire, seen in Fig.~\ref{fig:question}, the results are shown in Table.~\ref{tab:staff}, 5 people felt scared, 9 people felt funny looking, and 2 people commented that they were especially scared and wanted them to be taken off immediately.

\begin{figure}[!b]
    \centering
    \includegraphics[width=.5\textwidth]{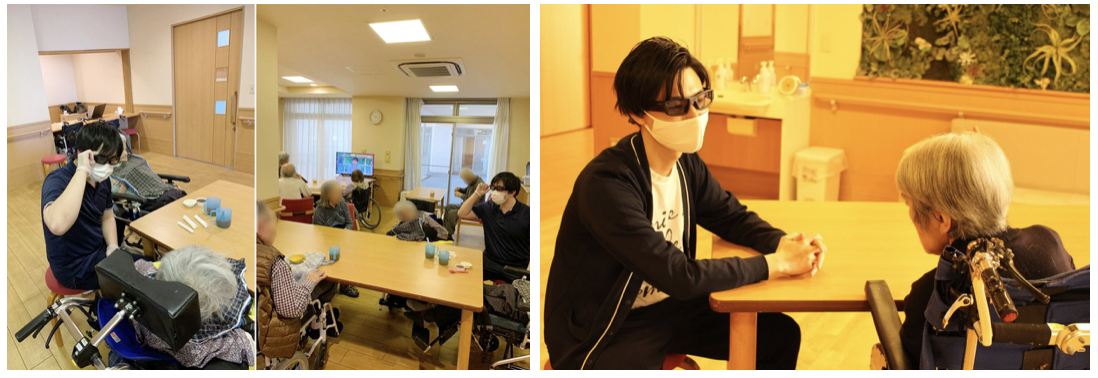}
    \caption{Questionnaire for caregivers in elderly care homes on spot.}
    \label{fig:question}
\end{figure}

Six caregivers participated in the experiments, the average rating of caregivers is 3.6, as shown in Table.~\ref{tab:staff-resident}.

% There are 6 caregivers and 20 residents join the experiments, the average rating of caregivers is 3.6, and 2.2 for the residents, shown in Table.~\ref{tab:staff-resident}.

\subsection{Analysis}
Our facial recognition achieved a high level of accuracy with a two-stage process incorporating face detection and facial recognition. However, facial detection relies on an external library~\cite{zhang2016joint} and this may lead to the system being unable to obtain facial features of some residents. In the future, we will consider improving facial detection for elderly people’s faces and detection in different posts, such as when they are lying in bed.

\textbf{Long-term care facilities}.
% It is confirmed that the accuracy of face recognition is high in the practical usage scenario, such as grasping the situation of residents in the corridor or dining room. In addition, by using this system, such as excretion information and body temperature could be grasped quickly hands-free, confirming the usefulness of our system. In the future, we will achieve more experiments with the change of place and invite more people to join in. 
It is confirmed that the accuracy of face recognition is high in the practical use scenario, such as ascertaining the location of residents in the corridor or dining room. In addition, by using the smart glasses, such information as that on excretion and body temperature could be obtained quickly hands-free, confirming the usefulness of our system. Due to the COVID-19 pandemic, we could not directly evaluate the questionnaire of the elderly residents. In the future, we will conduct more experiments with the change of location and invite more people to participate. 

\textbf{Privacy concern}.
An important factor of the SHECS system is the concerns for privacy when the caregivers review the residents' information. Residents may be concerned about information being displayed and read aloud on AR glasses during care. However, the truth is that the information displayed can only be seen by the caregiver, who retrieves it from the local server, and the voice read aloud is played by bone conduction headphones, which are also harder for people around the caregiver to hear. So the residents' individual information is safe.

\textbf{Extension of usage}.
We believe that the technology of the SHECS system can be used not only in elderly care but also in other fields. For example, in a nursery school, information about each child and their body temperature measured in the morning could be immediately detected, which could improve the efficiency of information sharing among the caregivers and understanding the health status of the children. With children sometimes hugging or falling, the handsfree system is effective in allowing both hands to be used. We think it can also be used to keep track of the names of kids' parents.% as well as the children.
% We believe that the technology of the SHECS system can be used not only in elderly care but also in other fields. For example, in a nursery school, information about each child such as the body temperature in the morning could be immediately detected, which could improve the efficiency of information sharing among the working staff and master the health status of the children. With children sometimes hugging or falling, the hands-free system allows the working staff to use both hands-frees or keep track of the names of their parents.

\section{Conclusion}
In order to instantly master the condition of elderly people in elderly care homes, it is necessary to promote the management of the care support system, such as providing permanent care for the elderly residents. In this paper, we develop a smart care system in conjunction with facial recognition technology and speech synthesis technology on smart AR glasses, in which the residents' information is provided by displaying part of the information and reading out the complete information out.

The emergence of the SHECS system is motivated by the requirements of both new and old elderly residents. This system promotes communication especially between caregivers and new residents so that they get to know each other beforehand. In the future, we will continue to utilize the power of communication and work with partners to conduct research that leads to solutions to social issues, with the aim of achieving sustainable growth and the development of society. Moreover, We will achieve experiments with more influence factors such as the the distance between caregivers and residents.

% To instantly obtain the health condition of elderly people in elderly care homes and provide long-term care services smoothly with information and communications technology (ICT), in this paper, we utilized AR smart glasses in conjunction with face recognition technology and speech synthesis technology to develop a system that provides resident identification and information presentation to caregiver in a hands-free manner.

% The “hands-free elderly care work support system” was born out of the need to be able to provide care to new residents as if they were already familiar to us. We will continue to utilize the power of communication and work with partners to conduct research that leads to solutions to social issues, with the aim of achieving sustainable growth and the development of society.

\section*{Acknowledgment}
The authors would like to thank Zenkokai and Vuzix Corporation for providing the venue for experiments and for supporting software implementation on smart glasses, respectively. We also express our sincere thanks to the experts from Systemsoft Inc: Zhiguang Zhou, Zheshuang Lyu, Wataru Nishioka, Megumi Komiya for their great support and advice on our proposed method and its optimization. 

\bibliographystyle{IEEEtran}
\bibliography{sample-base}

% that's all folks
\end{document}